\documentclass[11pt,a4paper]{article}

\usepackage{amsmath,amssymb,amsthm}

\usepackage{times}

\def\qed{$\Box$}

\newcommand\comment[1]{}

\newtheorem{theorem}{Theorem}
\newtheorem{proposition}[theorem]{Proposition}
\newtheorem{lem}[theorem]{Lemma}
\newtheorem{corollary}[theorem]{Corollary}
\newtheorem{Definition}[theorem]{Definition}

\newtheorem*{rema}{Remark}

\def\ra{\rangle}

\newcommand{\be}{\begin{eqnarray}}
\newcommand{\ee}{\end{eqnarray}}

\renewcommand{\epsilon}{\varepsilon}

\begin{document}

\title{\Large \bf Permutation groups, minimal degrees and quantum computing}

\author{
Julia Kempe\\
CNRS \& LRI, Universit\'e Paris-Sud\\
91405 Orsay Cedex, France\\
\and L{\'a}szl{\'o} Pyber\\
Mathematical Institute of the
Hungarian Academy of Sciences\\
P.O. Box 127, Budapest, Hungary H-1364
 \and
Aner Shalev\\
Institute of Mathematics\\
The Hebrew University\\
Jerusalem 91904, Israel\\
}

\date{\today}

\maketitle

\begin{abstract}
We study permutation groups of given minimal degree without the classical
primitivity assumption. We provide sharp upper bounds on the order of a
permutation group $H \le S_n$ of minimal degree $m$ and on the number of
its elements of any given support. These results contribute to the
foundations of a non-commutative coding theory.

A main application of our results concerns the Hidden
Subgroup Problem for $S_n$ in Quantum Computing. We completely characterize
the hidden subgroups of $S_n$ that can be distinguished from identity with
weak Quantum Fourier Sampling, showing these are exactly the subgroups with
bounded
minimal degree. This implies that the weak standard method for $S_n$ has no
advantage whatsoever over classical exhaustive search.
\end{abstract}

\newpage

\section{Introduction}

Let $S_n$ denote the symmetric group on $\{1,\ldots,n\}$. For a permutation
$h \in S_n$ define its support $supp(h)$ by
$$supp(h) = \{i \in \{1,\ldots ,n\}: h(i) \neq i\}.$$
The {\it minimal degree} $m(H)$ of a permutation group $1 \ne H \le S_n$ is
defined to be the minimal number of points moved by a non-identity element of
$H$. In other words,
\[
m(H) = \min \{ |supp(h)|: 1 \ne h \in H \}.
\]
This notion goes back to the 19th century, and plays an important role in the
theory of finite permutation groups since the days of Jordan
\cite{Jordan:73a,Jordan:75a}. Particular attention was given to the minimal
degree of {\em primitive} permutation groups. Recall that a permutation group
is called primitive if it is transitive and doesn't preserve a non-trivial
block system. Let $H <S_n$ be a primitive permutation group not containing
$A_n$. Jordan proved that $m(H)$ goes to infinity as $n$ goes to infinity.
Babai \cite{Babai:81a} showed that under the above conditions we actually have
that
$$m(H) \geq \frac{\sqrt{n}-1}{2}.$$
This result is essentially best possible. However, if we exclude certain
primitive groups and use the Classification of Finite Simple Groups (CFSG),
sharper bounds can be obtained. Indeed, it was shown by Liebeck and Saxl in
\cite{LS91} that $m(H) \geq n/3$ with a given list of exceptions. This lower
bound was improved by Guralnick and Magaard in \cite{GM98} to $n/2$
(with prescribed exceptions).
See also
Cameron \cite{Cameron:82a} for the impact of the Classification on the theory
of finite permutation groups and primitive groups in particular.

In spite of considerable progress in the study of the minimal degree of
primitive groups, much less is known in the non-primitive case. One of the
purposes of this paper is to study permutation groups of given minimal degree
without assuming primitivity or even transitivity.

A basic question in this field is: how large can a permutation group $H$ of
degree $n$ and minimal degree $m$ be? An easy classical upper bound is $|H|
\leq n^{n-m+1}$. Indeed, this follows from the fact that a permutation $h \in
H$ is uniquely determined by its action on $\{ 1, \ldots , n-m+1 \}$.

Better bounds were given by Liebeck \cite{Li1,Li2} under the assumption
that $H$ is transitive.
Our first result extends Liebeck's theorem to arbitrary permutation groups.

\setcounter{theorem}{0}
\renewcommand{\thetheorem}{\Alph{theorem}}

\begin{theorem}
\label{thm:A} Let $H \le S_n$ be a permutation group with minimal degree
$m = m(H)$.

{\rm 1)} If $m \leq \log_2 n$, then $|H| \leq n^{10 n/m}$.

 {\rm 2)} If $m \geq \log_2 n$, then $|H| \leq 2^{10n}$.
\end{theorem}

Theorem \ref{thm:A} is essentially best possible. For example, consider the
group $H=S_{2n/m} <S_n$ acting on $2n/m$ blocks of size $m/2$.
Then the minimal degree of $H$ is $m$ and $|H|=(2n/m)!$ which is of the
form $n^{(2-o(1))n/m}$ when $m \leq \log_2 n$. Up to a constant in the
exponent, this shows that part (1) of Theorem \ref{thm:A} is tight.

Note also that if $H \le S_n$ is transitive of minimal
degree $m$ and base size $b$, then $bm \ge n$ (see e.g. \cite{DM}, p. 80),
and this implies $|H| \ge 2^b \ge 2^{n/m}$.
\medskip

Subgroups of $S_n$ of given minimal degree $m$ can be regarded as
non-commutative analogues of linear codes with minimal distance $m$. Recall
that in coding theory \cite{MacWilliams} a fundamental question is: how large
can a subspace of $GF(q)^n$ with minimal distance $m$ be? Replacing the
Abelian group $GF(q)^n$ by the symmetric group $S_n$ we may ask a similar
question in this context. Theorem \ref{thm:A} provides a rather sharp
answer.

Note that any binary linear code inside $GF(2)^{n/2}$ can be embedded
naturally
as a subgroup of $S_2^{n/2}<S_n$. Thus classical coding theory provides a rich
source of constructions of permutation groups of large minimal degree. In
particular the (obvious) Gilbert-Varshamov lower bound (\cite{handbookcomb} p.
781, remark after Thm. 3.5) applied to linear codes produces exponentially
large elementary Abelian permutation groups with large minimal degree, e.g.
$m>n/8$. This demonstrates the tightness of part (2) of Theorem \ref{thm:A},
even when $m$ is very large.
\medskip

Another classical question in coding theory is the study of the
{\it weight distribution}, namely counting elements of weight $k$ in
a code with minimal distance $m$.
The analogous question for permutation groups is counting
the number of elements of support $k$ in a permutation group of minimal
degree $m$. Given a permutation group $H \le S_n$ define
\[
H_k = \{ h \in H: |supp(h)|=k \},
\]
the subset of elements of support $k$ in $H$. In our second result, which
is the most technically demanding, we bound the size of $H_k$.

\begin{theorem}\label{th:size}
There exists absolute constants $b, \epsilon >0$ such that if a subgroup
$H \le S_n$ has minimal degree $m \geq b$ then
\begin{equation*}
|H_k|\leq n^{-\varepsilon m} \binom {n}{k}^{\frac{1}{2}} (k!)^{\frac{1}{4}}.
\end{equation*}
\end{theorem}

The theorem has an interesting consequence for the number of elements of
minimal support.
If $k=m \le n^{2 \varepsilon}$ then $(k!)^{1/4} \le n^{\varepsilon m/2}$
and this implies
\[
|H_m| \leq n^{-\varepsilon m/2} \binom {n}{m}^{1/2}.
\]

This upper bound is essentially tight. To show this we use some results from
coding theory and the above embedding of binary codes in $S_n$. Consider the
well known Goppa code \cite{Goppa:70a} and the estimates for the number of code
words of minimal weight \cite{Litsyn:97a}. For a binary Goppa code over
$GF(2)^{n/2}$, in the regime of small $t$ ($t \ll \sqrt{\log n}$), the number
of code words of minimum weight $2t+1$ is roughly (up to a constant factor)
$$\binom{n/2}{2t+1} (\frac{n}{2})^{-t}.$$
Embedding this code into $S_n$ as above, we obtain a subgroup $H<S_n$ of
minimal degree $m=4t+2$ satisfying
$$|H_m| \geq c n^{-m/4} \binom {n}{m}^{\frac{1}{2}} $$
for some constant $c>0$. This demonstrates the tightness of Theorem
\ref{th:size} in the regime of small $m$.

\medskip

A main motivation behind Theorem \ref{th:size}, besides the study of
weight distributions of non-commutative codes, comes from Quantum Computing.
A central problem in Quantum Computing is the Hidden Subgroup Problem
(HSP), which we state below.
Let $G$ be a finite group and $H \le G$ a subgroup.
Given a function $f: G \rightarrow S$ that is constant on
(left)-cosets $gH$ of $H$ and takes different values for different cosets,
determine a set of generators for $H$. The decision version of this problem
is to determine whether there is a non-identity hidden subgroup or not.

Note that given $g \in G$ we have $g \in H$ if and only if $f(g) = f(1)$.
Using classical search we may therefore perform membership tests,
and once we find a non-identity element $g \in H$ we may conclude
that $H \ne 1$. However, the aim is to decide whether or not $H = 1$
in polynomial time, namely after $(\log{|G|})^c$ steps.
Complete enumeration over the elements $g \in G$ is therefore not
efficient. The question is whether a quantum computer can solve the
HSP efficiently (giving the correct answer in polynomial time
with a very high probability).

The Hidden Subgroup Problem plays a central role in Quantum Computing. Nearly
all quantum algorithms which significantly improve the known classical
algorithms, like factoring and discrete log, solve the Abelian version of this
problem by the so called standard method of Quantum Fourier Sampling. One of
the most important questions is whether the standard method can efficiently
solve the {\em non-Abelian} HSP, especially for the symmetric group $G=S_n$.
This latter case in particular would yield a quantum algorithm for the Graph
Isomorphism Problem, for which no efficient classical algorithm is known. For
more details on Quantum Computing, the HSP, and the standard method see
Section \ref{sec:quantum}.

To state our main quantum-theoretic application in a precise
mathematical way we need some notation. Given a finite group
$G$ let $Irr(G)$ denote the set of (complex) irreducible representations
of $G$ (up to equivalence). For $\rho \in Irr G$ let $d_{\rho}$ denote
its dimension and $\chi_{\rho}$ its character.

Given a subgroup $H \le G$, define
\begin{equation}\label{DH}
D_H=\frac{1}{|G|} \sum_{\rho \in Irr G} d_\rho
|\sum_{h \in H, h \neq 1} \chi_\rho(h)|.
\end{equation}

Roughly speaking, $D_H$ measures the $L_1$-distance between a
(non-commutative) Fourier transform of the characteristic
function of $H$ and that of the characteristic function of
the identity.

We say that a subgroup $H \le G$ is {\it distinguishable}
if
\[
D_H \ge (\log{|G|})^{-c}
\]
for some constant $c$. Of course this is an asymptotic notion,
where we think of $G$ as ranging over an infinite family of groups,
whereas the constant $c$ does not depend on $G$. Here we focus on
the case $G = S_n$, where distinguishability is equivalent to
$D_H \ge n^{-c}$.
Distinguishable subgroups $H$ are those which can be distinguished
from 1 using the so called weak standard method (see the next section
for more details).

The main application of this paper to Quantum Computing, which
relies heavily on Theorem \ref{th:size} above, is the following.

\setcounter{theorem}{2}
\renewcommand{\thetheorem}{\Alph{theorem}}
\begin{theorem}\label{thm:c}
Let $H\leq S_n$  be a subgroup. If $H$ is distinguishable, then it has a
bounded minimal degree. Moreover, if $D_H \ge n^{-c}$, then $m(H) \le g(c)$,
where $g(x) = ax+b$ is some fixed {\it linear} function.
\end{theorem}

Thus all subgroups of unbounded minimal degree are indistinguishable,
which opens up a huge spectrum of examples and constructions.
The only case previously known in the literature of an indistinguishable
subgroup of $S_n$ is that of a subgroup of order $2$ generated by a
fixed point free involution or by a product of transpositions of large
support \cite{Hallgren:00a,Grigni:01a}.
Obviously $m(H)$ is unbounded for such subgroups $H$, so
its indistinguishablity is an immediate consequence of the above theorem.

In an
extended abstract \cite{KS:05} a subset of the authors of this paper have
proved a weaker version of Theorem \ref{thm:c} (for primitive subgroups and
subgroups of polynomial size) and have conjectured that it holds in full
generality. This paper proves the conjecture.

It is intriguing that much larger subgroups are also indistinguishable.
Indeed take $H = S_{2n/m} < S_n$, the subgroup constructed following
Theorem A. If $m = m(H)$ tends to infinity arbitrarily slowly, then $H$
is indistinguishable and
$|H| \ge (n!)^{\epsilon (n)}$ where $\epsilon (n)$ tends to $0$ arbitrarily
slowly. In particular, the size of indistinguishable subgroups of $S_n$ can be
super-exponential in $n$.

However, if $\epsilon > 0$ is fixed, and $|H| \ge (n!)^{\epsilon}$,
then it follows from Theorem A that the minimal degree of $H$
is bounded. Enumerating over elements of $S_n$ of bounded support
(their number is bounded by a polynomial in $n$) we deduce that such
a subgroup $H$ can be distinguished from 1 using classical search.

It follows from the two paragraphs above that {\em all} subgroups $H \le S_n$
of size $\ge N$ can be distinguished from 1 using the weak standard method
(together with classical search) if and only if $N \ge (n!)^{\epsilon}$ where
$\epsilon$ is bounded away from zero.

Theorem C has rather grave consequences. Indeed, if $H$ is distinguishable
then it has an element of bounded support, and this can be detected
(as above) after polynomially many membership tests (when we enumerate
the permutations in $S_n$ according to their support).

\begin{corollary}\label{cor:D}
Any subgroup $H \leq S_n$ which is distinguishable can already be
distinguished from $1$ using classical search.
\end{corollary}

Thus Theorem C provides a
complete characterisation of hidden subgroups $H\leq S_n$
which can be distinguished from 1 using the weak standard method
and classical search: these are precisely the subgroups of bounded
minimal degree.

It is intriguing that the old classical notion of minimal degree,
which is central in the theory of finite permutation groups,
plays a role in the context of quantum computing.
The Classification of Finite Simple Groups (CFSG) is also used in an
essential way in some parts of this work.

\medskip

Some words on the structure of this paper. In Section \ref{sec:quantum} we
provide background on quantum computing, the Hidden Subgroup Problem, and the
standard method of Quantum Fourier Sampling. Section \ref{s:3} deals with
arbitrary finite groups $G$ and their subgroups $H$. Using character-theoretic
methods we give upper and lower bounds on the $L_1$-distance $D_H$ introduced
above. We then characterize distinguishable subgroups of polylogarithmic size.
In Section \ref{s:4} we focus on the case $G = S_n$. We prove there (relying on
CFSG and other tools) that any primitive subgroup $H < S_n$ not containing
$A_n$ is indistinguishable. We also show how to deduce Theorem C from Theorem
B. Theorem A is proved in Section \ref{s:5}. Section \ref{s:6}, which is the
longest in this paper, is devoted to the proof of Theorem B. This proof applies
Theorem A as well as results on primitive groups obtained in Section \ref{s:4}.

\bigskip
\bigskip

\section{Quantum Computing}\label{sec:quantum}


In the last decade quantum computation has provided us with powerful tools to
solve problems not known to be classically efficiently solvable, like factoring
and discrete log \cite{Shor:94a}. Nearly all the problems in which a quantum
computer excels more than quadratically with respect to its classical
counterpart can be cast into the framework of the Hidden Subgroup Problem
(HSP).  Let $G$ be a finite group and $H \le G$ a subgroup. Given a function
$f: G \rightarrow S$ that is constant on (left)-cosets $gH$ of $H$ and takes
different values for different cosets, determine a set of generators for $H$.
The decision version of this problem is to determine whether there is a
non-identity hidden subgroup or not.

The reason that quantum computers seem to provide a speed-up for this type of
problem is that it is possible to implement the Fourier transform over certain
groups {\em efficiently} on a quantum computer. This in turn allows to sample
the Fourier components efficiently (this technique of Quantum Fourier Sampling
is referred to as the ``standard method''). In the case of Abelian groups $G$
(appearing in factoring and discrete log) the hidden subgroup can be
reconstructed with only a polynomial (in $\log |G|$) number of queries to the
function and a polynomial number of measurements (samplings in the Fourier
basis) and postprocessing steps.

We denote states of the vector space $\mathbb{C}[G]$, spanned by the group
elements, with a $|\cdot\ra$, as is standard in quantum computation (see e.g.
\cite{Nielsen:book} for more details).

\renewcommand{\thetheorem}{\arabic{theorem}}
\setcounter{theorem}{0}

\begin{Definition}
The Quantum Fourier Transform (QFT) over a group $G$ is the following unitary
transformation on $\mathbb{C}[G]$:
$$|g\ra \rightarrow \frac{1}{\sqrt{|G|}} \sum_{\rho,i,j} \sqrt{d_\rho} \rho(g)_{ij} |\rho,i,j\ra$$
where $\rho$ labels an irreducible representation of $G$, $d_\rho$ is its
dimension and $1 \leq i,j \leq d_\rho$. The $|\rho,i,j\ra$ span another basis
of $\mathbb{C}[G]$, the so called Fourier basis.
\end{Definition}
For many non-Abelian groups it is possible to implement the Fourier transform
on a quantum computer efficiently, and in particular explicit constructions
exist for the symmetric group $S_n$ \cite{Beals:97a}.

Addressing the HSP in the non-Abelian case is considered to be one of the most
important challenges at present in quantum computing.  A positive answer to the
question whether quantum computers can efficiently solve the Hidden Subgroup
Problem over non-Abelian groups would have several important implications for
the solution of problems in NP, which are neither known to be NP-complete nor
in P; and which are good candidates for a quantum speed-up. Among the most
prominent such problems is Graph Isomorphism, where the group in question is
the symmetric group. Hence it is very desirable to get a handle on the power of
Quantum Fourier Sampling (QFS) to solve the HSP for general groups.

\begin{Definition}
The {\it standard method} of Quantum Fourier Sampling is the following: The
state is initialised in a uniform superposition over all group elements; a
second register is initialised to $|0\ra$. Then the function $f$ is applied
reversibly over both registers (i.e. $f:|g\ra |0\ra \rightarrow |g\ra
|f(g)\ra$). The second register is measured, which puts the first register into
the superposition of a (left)-coset of $H$, i.e.  in the state
$|gH\ra:=\frac{1}{\sqrt{|H|}} \sum_{h \in H} |gh\ra$ for some random $g \in G$.
Finally the QFT over $G$ is performed, yielding the state
$$\frac{1}{\sqrt{|G||H|}} \sum_{\rho,i,j} \sqrt{d_\rho} \sum_{h \in H} \rho_{ij}(gh)|\rho,i,j\ra.$$
A basis measurement now gives $(\rho,i,j)$ with probability
$P_{gH}(\rho,i,j)=\frac{d_\rho}{|G||H|}|\sum_{h \in H}\rho_{ij}(gh)|^2.$
\end{Definition}
Since we do not know $g$ and $g$ is distributed uniformly, we sample
$(\rho,i,j)$ with probability $P_{H}=\frac{1}{|G|}\sum_{g} P_{gH}$. The {\it
strong} standard method samples both $\rho$ and its entries $i,j$.  In the {\it
weak} standard method  only the character $\chi_\rho$ is measured  (but not the
entries $i,j$, which are averaged over). In this case it is not hard to see
\cite{Hallgren:00a,Grigni:01a} that the probability to sample $\rho$ is
independent of the coset of $H$ we happen to land in. Hence the probability to
measure $\rho$ in the weak case is
$$P_H(\rho)=\frac{d_\rho}{|G|}\sum_{h \in H} \chi_\rho(h).$$
Note that from this expression it is clear that the weak standard method cannot
distinguish between conjugate subgroups \cite{Hallgren:00a}. Let $Irr(G)$ be
the set of irreducible characters of $G$. Then $P_H$ is a distribution on
$Irr(G)$. The strong standard method sometimes provides substantially more
information than its weak counterpart, and is indeed necessary to efficiently
solve the HSP in the case of groups like the Dihedral group
\cite{Ettinger:99a,Kuperberg:03a,Regev:04a} and other semidirect product groups
\cite{Moore:04a}. However (see below), for $S_n$ Grigni et al.
\cite{Grigni:01a} have shown that for a {\em random} basis the additional
information provided by the strong method is exponentially small except
possibly for very large subgroups.

An even more basic question is which hidden subgroups can be {\em distinguished
from the identity} via QFS with special attention to the symmetric group.
Distinguishing the trivial subgroup $\{e\}$ from a larger subgroup $H$
efficiently using the weak standard method is possible if and only if the $L_1$
distance $D_H$ between $P_{\{e\}}$ and $P_H$ is larger than some inverse
polynomial in $\log |G|$. The $L_1$ distance (also known as the total variation
distance) is given as
\[
D_H=\frac{1}{|G|} \sum_\rho d_\rho |\sum_{h \in H, h \neq 1} \chi_\rho(h)|.
\]
We say that $H$ is {\it distinguishable} (using the weak standard method) if
$D_H \ge (\log{|G|})^{-c}$ for some constant $c$, and {\it indistinguishable}
otherwise.

Several positive results on the power of QFS for the Hidden Subgroup Problem
have been obtained previously for groups that are in some ways ``close'' to
Abelian, like some semidirect products of Abelian groups
\cite{Ettinger:99a,Roetteler:98a,Kuperberg:03a,Regev:04a,Moore:04a}, in
particular the Dihedral group; Hamiltonian groups \cite{Hallgren:00a}, groups
with small commutator groups \cite{Ivanyos:01a} and solvable groups of constant
exponent and constant length derived series \cite{Friedl:03a}. Often in these
cases the irreducible representations are known and can be analysed.
For instance the Dihedral group $D_n$, the first non-Abelian group to be
analysed in this context by Ettinger and Hoyer \cite{Ettinger:99a}, is
``nearly'' Abelian in the sense that all of its irreducible representations
have degree at most two.  Indeed hidden reflections of $D_n$ can be
distinguished from the identity with only polynomial Quantum Fourier Samplings,
similar to the Abelian case (where all irreducible representations are
one-dimensional). Note, however, that the computational version of the HSP
seems much harder: even though a polynomial number of samples suffice to {\em
distinguish} hidden reflections {\em information theoretically}, no efficient
reconstruction procedure is known.

The holy grail of the field is the symmetric group $S_n$, which seems much
harder to analyse, partly because to this day there is still only partial
explicit knowledge about its irreducible representations and character values
\cite{Sagan:book}, because most of its subgroups are far from normal (have many
conjugate subgroups), because most of its irreducible representations have very
large dimension ($2^{\Theta (n \log n)}$) and the number of different
irreducible representations is an exponentially small fraction of the size of
the group, to name just some of the difficulties. The structure of
distinguishable versus indistinguishable subgroups of $S_n$ has remained open.

The following results have been obtained for the HSP over the symmetric group:
The group $S_n$ being non-Abelian, Quantum Fourier Sampling gives a
distribution on both the characters and the entries of the corresponding matrix
representations. Grigni et al. \cite{Grigni:01a} show that sampling the row
index in the strong standard method provides no additional information. They
also show that the additional information provided by the strong method in a
{\em random} basis scales with $\sqrt[3]{|H|^2 k(G)/|G|}$ where $k(G)$ is the
number of conjugacy classes of the group $G$ and $|H|$ the size of the hidden
subgroup. Both Hallgren et al. and Grigni et al. \cite{Hallgren:00a,Grigni:01a}
show that hidden subgroups of $S_n$ of size $|H|=2$, generated by involutions
with large support, cannot be distinguished from identity; exactly the task
that needs to be solved for Graph Automorphism. Recently, Moore et al. have
essentially shown that the {\em strong} standard method cannot distinguish the
subgroup generated by a fixed point free involution from identity
\cite{Moore:05a}. Moreover, even a generalization of the strong standard method
to $O(n \log n)$ instances of Quantum Fourier Sampling does not allow to
distinguish the above subgroup from $1$ \cite{Hallgren:06a}. No results are
known for other subgroups of $S_n$.


In this work various classical as well as modern parts of the theory of
permutation groups are applied for the first time in the context of quantum
computing. In our applications to the hidden subgroup problem, we focus on the
{\em weak} form of the standard method, since the strong form with random
choices of basis does not provide any non-negligible additional information for
the symmetric group and the subgroups we consider \cite{Grigni:01a}. It remains
to be seen whether judicious choices of basis for each irreducible matrix
representation can give more information in the case where random choices don't
help; but to our knowledge no such examples have been found and in fact recent
results of Moore et al. \cite{Moore:05a} show that in the case of fixed point
free involutions no such good basis exists.

Theorem \ref{thm:c} and Corollary \ref{cor:D} above provide a complete
characterization of subgroups which can be distinguished from $1$ using the
weak standard method (together with classical exhaustive search). Indeed, these
are exactly the subgroups of $S_n$ with bounded minimal degree. For instance we
cannot distinguish a group generated by a cycle of unbounded length or an
involution with unbounded number of transpositions (implying the result in
\cite{Hallgren:00a,Grigni:01a}).

This also has implications for the Graph Isomorphism (GI) problem. Recall that
to solve GI for two graphs $G_1,G_2$, it suffices to distinguish a hidden
subgroup of the automorphism group $Aut(G_1 \cup G_2)$ of the form $H_1 \times
H_2$ (not $G_1 \simeq G_2$), where $H_i=Aut(G_i)$, from a subgroup of the form
$H \cup \sigma H$ ($G_1 \simeq G_2$), where $H=H_1 \times H_2$ and $\sigma$
maps $G_1$ to $G_2$. Our results imply that we cannot distinguish each of the
two possible cases from identity, and hence (using the triangle inequality) we
cannot distinguish them from each other unless Aut($G_i$) contains an element
of bounded support. Thus weak QFS provides no advantage here.

\bigskip
\bigskip

\section{Arbitrary groups}\label{s:3}

In this section we discuss results for arbitrary finite groups $G$.
Our starting point
is a general result providing both upper and lower bounds on the total
variation distance $D_H$ in terms of the same group theoretic data.
While the definition of $D_H$ involves character degrees and values,
which are hard to compute, our bounds below involve sizes of conjugacy
classes, and their intersections with the hidden subgroup.

We need some group theoretic notation. For $h \in G$ we let $h^G$ denote the
conjugacy class of $h$ in $G$. Let $C_1, \ldots , C_k$ denote the non-identity
conjugacy classes of $G$. For an irreducible character $\chi_{\rho} \in Irr(G)$
we let $\chi_{\rho}(C_i)$ denote the common value of $\chi_{\rho}(x)$ for
elements $x \in C_i$.

\renewcommand{\thetheorem}{\arabic{theorem}}
\setcounter{theorem}{0}

\begin{proposition}\label{Th:main}
Let $H<G$. Then
\begin{align*}
&1. \;\;\; \sum_{i=1}^k |C_i \cap H|^2 |H|^{-1}|C_i|^{-1}< D_H\\
&2. \;\;\; D_H \le \sum_{i=1}^k |C_i \cap H||C_i|^{-\frac{1}{2}}=
\sum_{1 \ne h \in H} |h^G|^{-1/2}.
\end{align*}
\end{proposition}

Applying the upper bound with $|H|=2$ gives the result obtained previously by
Hallgren et al. and Grigni et al. \cite{Hallgren:00a,Grigni:01a}. No lower
bounds seem to exist in the literature. This result has a wide range of
applications. For example, it enables us to characterise distinguishable
subgroups $H \le G$ of polylogarithmic order
(see Theorem \ref{Th:polygeneral} below).
\medskip

\noindent {\it Proof of Proposition \ref{Th:main}.} For each irreducible
representation $\rho$ of $G$ we have
\begin{align*}
|\sum_{h \in H, h \neq 1} \chi_\rho(h)| & \le \sum_{h \in H, h \neq 1}
|\chi_\rho(h)|   \le \sum_{h \in H, h \neq 1} d_\rho < |H|d_{\rho}.
\end{align*}
Hence
$d_{\rho} > |H|^{-1}|\sum_{h \in H, h \neq 1} \chi_\rho(h)|$.
Substituting this in (\ref{DH}) we obtain
\[
D_H > \frac{1}{|G||H|}
\sum_{\rho} |\sum_{h \in H, h\neq 1} \chi_\rho(h)|^2.
\]
Note that $\chi_\rho(h) = \chi_{\rho}(C_i)$
if $h \in H \cap C_i$. This yields
$\sum_{h \in H, h \neq 1} \chi_\rho(h) =
\sum_{i=1}^k |H \cap C_i| \chi_\rho(C_i)$,
and so
\[
D_H > \frac{1}{|G||H|}\sum_{\rho}
|\sum_{i=1}^k |H \cap C_i| \chi_\rho(C_i)|^2.
\]
Now,
\begin{align*}
|\sum_{i=1}^k |H \cap C_i| \chi_\rho(C_i)|^2 = \sum_{i=1}^k |H \cap C_i|^2
|\chi_\rho(C_i)|^2 + \sum_{i \ne j} |H \cap C_i||H \cap C_j| \chi_\rho(C_i)
{\bar{\chi}_{\rho}}(C_j).
\end{align*}

Using the generalised orthogonality relations we observe that
\[
\sum_{\rho} \sum_{i=1}^k |H \cap C_i|^2 |\chi_\rho(C_i)|^2
= \sum_{i=1}^k |H \cap C_i|^2 |G|/|C_i|,
\]
and
\[
\sum_{\rho} \sum_{i \ne j} |H \cap C_i||H \cap C_j|
\chi_\rho(C_i) {\bar{\chi}_{\rho}}(C_j) = 0.
\]
It follows that
\begin{align*}
D_H &> \frac{1}{|G||H|} \sum_{i=1}^k |H \cap C_i|^2 |G|/|C_i|  = \sum_{i=1}^k
|H \cap C_i|^2 |H|^{-1}|C_i|^{-1}.
\end{align*}
This completes the proof of the lower bound.

To prove the upper bound, write
\begin{align}\label{Eq:upperDH}
D_H |G|&=\sum_{\rho} d_\rho |\sum_{h \in H, h\neq 1} \chi_\rho(h)| \le
\sum_{\rho} d_\rho \sum_{h \in H, h\neq 1} |\chi_\rho(h)| = \sum_{h \in H,
h\neq 1} \sum_{\rho} d_{\rho} |\chi_\rho(h)|.
\end{align}

Fix $h \in H$ and choose $i$ such that $h \in C_i$.
Using the Cauchy-Schwarz inequality we obtain
\[
\sum_{\rho} d_{\rho} |\chi_\rho(h)| \le
(\sum_{\rho} d_{\rho}^2)^{1/2} (\sum_{\rho} |\chi_{\rho}(h)|^2)^{1/2},
\]
giving (using the orthogonality relations)
\[
\sum_{\rho} d_{\rho} |\chi_\rho(h)| \le |G|^{1/2}(|G|/|C_i|)^{1/2} =
|G||C_i|^{-1/2}.
\]
Summing over non-identity elements $h \in H$, and observing that
the upper bound above occurs $|H \cap C_i|$ times, we obtain
\[
\sum_{h \in H, h \ne e} \sum_{\rho} d_{\rho} |\chi_\rho(h)| \le
\sum_{i=1}^k  |H \cap C_i||G||C_i|^{-1/2}.
\]
Combining this with
(\ref{Eq:upperDH})
we obtain
\[
D_H \le \sum_{i=1}^k |H \cap C_i| |C_i|^{-1/2},
\]
as required.
\qed
\medskip

The following is an immediate consequence of Proposition \ref{Th:main}.

\begin{corollary} \label{corr:1}
Let $C_{min}$ denote a non-identity conjugacy class of minimal
size intersecting $H$ non-trivially. Then we have
\[
|H|^{-1}|C_{min}|^{-1} < D_H \le (|H|-1)|C_{min}|^{-1/2}.
\]
\end{corollary}

We can now prove the main result of this section, characterising
distinguishable subgroups of polylogarithmic order in an arbitrary
group $G$.

\begin{theorem}\label{Th:polygeneral}
Suppose $|H| \le (\log{|G|})^c$ for some constant $c$.
Then $H$ is distinguishable if and only if $H$ has a non-identity element
$h$ such that $|h^G| \le (\log{|G|})^{c'}$ for some constant $c'$.
\end{theorem}

\noindent
{\it Proof.}

Suppose first that $H$ is distinguishable, namely
$D_H \ge (\log{|G|})^{-b}$ for some constant $b$.
Then the upper bound in the above corollary shows that
\[
|H||C_{min}|^{-1/2} \ge  (\log{|G|})^{-b},
\]
so
\[
|C_{min}| \le  |H|^2 (\log{|G|})^{2b}) \le (\log{|G|})^{2(b+c)}.
\]
In the other direction, suppose $|C_{min}| \le  (\log{|G|})^{b}$.
Then the lower bound in the corollary above gives
\[
D_H > |H|^{-1}  (\log{|G|})^{-b} \ge  (\log{|G|})^{-(b+c)}.
\]
The result follows.
\qed
\medskip

\bigskip
\bigskip

\section{Symmetric groups} \label{s:4}

Let us now focus on the case $G = S_n$. In this section we first prove some
preliminary results related to distinguishability of subgroups of $S_n$. Some
of these results play a role in the proof of Theorem
\ref{th:size}. We also deduce
 Theorem \ref{thm:c} from Theorem \ref{th:size}.

\begin{proposition} \label{Th:poly}
Let $H \le S_n$ with $|H| \le n^c$ for some constant $c$. Then $H$ is
distinguishable if and only if its minimal degree $m(H)$ is bounded.
\end{proposition}

\noindent {\it Proof.} Let $g \in S_n$ with $supp(g)=k$. Then it is
straightforward to verify that $\binom {n}{k} \le |g^{S_n}| \le n^k$.
As a consequence we see that a conjugacy class $C$ in $S_n$ has polynomial
order if
and only if it consists of elements of bounded support.
This observation, when
combined with Theorem \ref{Th:polygeneral}, completes the proof. \qed
\medskip

Our next result concerns primitive subgroups. Primitive permutation groups are
considered the building blocks of finite permutation groups in general, and
were extensively studied over the past 130 years. We note that if $H \le S_n$
is primitive and $H \ne A_n, S_n$ then Babai showed that $|H| \le
n^{4\sqrt{n}\log{n}}$. Using the Classification of Finite Simple Groups the
latter bound can be somewhat improved  to $|H| \le 2n^{\sqrt{n}}$, which is
essentially best possible \cite{Cameron:82a}; in particular the order of $H$
can be much more than polynomial, and so Proposition \ref{Th:poly} above does
not apply.

However, we obtain the following somewhat surprising general result:

\begin{theorem}\label{Th:primitive}
Let $H \ne A_n, S_n$ be a primitive subgroup. Then $H$ is indistinguishable.
Moreover, there is an absolute constant $\epsilon>0$ such that
\[
D_H \le n^{ -\epsilon \sqrt{n} }.
\]
\end{theorem}

This theorem follows immediately from the two technical lemmas below,
which are based on counting elements of given support in permutation
groups $H$.
Recall that for $H \le S_n$ we set
\[
H_k = \{ h \in H: |supp(h)|=k \}.
\]

\begin{lem} \label{Prop}
Let $H \le  S_n$ be a subgroup. Suppose
that, for each $k \le n$, we have
\[
|H_k| \le n^{(1/6 - \epsilon)k}.
\]
where $\epsilon > 0$ is some fixed constant.
Then, if $n$ is large enough (given $\epsilon$) we have
\[
D_H \le 2n^{-\delta m(H)},
\]
where $\delta = \epsilon/2$.
In particular, if the minimal degree $m(H)$ is unbounded,
then $H$ is is indistinguishable.
\end{lem}
\paragraph{Proof:}
Apply the upper bound of Proposition \ref{Th:main}, written in the form
\[
D_H \le \sum_{1 \ne h \in H} |h^G|^{-1/2}.
\]
To evaluate this sum we use a result from \cite{Liebeck:01a},
showing that, for $G=S_n$ and $h \in G$ of support $k$
we have $|h^G| > n^{ak}$ for any real $a < 1/3$ and $n$
large enough (given $a$). Using this we obtain
\[
D_H < \sum_{k \ge m(H)} |H_k|n^{-bk},
\]
for any real number $b < 1/6$ and sufficiently large $n$.
Let $\delta = \epsilon/2$, $b = 1/6 - \delta$, and $m = m(H)$.
Then the upper bound on $|H_k|$ yields
\[
D_H < \sum_{k \ge m} n^{(1/6-\epsilon)k}n^{-(1/6-\delta)k} =
\sum_{k \ge m} n^{-\delta k} \le 2n^{- \delta m}.
\]
This proves the first assertion. Assuming $m=m(H)$ is unbounded, we see that
$D_H$ is smaller than any fixed negative power of $n$, and so $H$ is
indistinguishable. \qed
\medskip

\begin{lem}\label{lem:oneseventh}
Let $H<S_n$ be primitive and $H \neq A_n,S_n$. Then for sufficiently large $n$
and for all $k$ we have $|H_k| \leq n^{\frac{k}{7}}$.
\end{lem}
\paragraph{Proof:}
We use Babai's lower bound on the minimal degree of primitive subgroups $H \ne
A_n, S_n$ \cite{Babai:81a}, showing that
\begin{equation}\label{Eq:B}
m(H) \ge (\sqrt{n}-1)/2.
\end{equation}
Furthermore, we apply a theorem of Cameron \cite{Cameron:82a}
(which in turns relies on the Classification of Finite Simple Groups)
describing all primitive groups of `large' order. In particular it
follows from that description that, for all large $n$, and for a primitive
subgroup $H \ne A_n,S_n$, either

(i) $|H| \le n^{c n^{1/3}}$, or

(ii) $n = \binom{l}{2}$ for some $l$, and $H \le S_l$ acting
on $2$-subsets of $\{1, \ldots , l \}$, or

(iii) $n=l^2$ for some $l$, and $H \le S_l \wr S_2$ acting on
$\{ 1, \ldots , l \}^2$ in the so called product action.

We claim that for all large $n$ and for all $k$ we have $|H_k| \le n^{k/7}$. To
show this it suffices to consider $k \ge (\sqrt{n}-1)/2$, otherwise $|H_k|=0$
by (\ref{Eq:B}). Now, if $H$ satisfies condition (i) above then the claim
follows trivially using $|H_k| \le |H|$. So it remains to consider groups $H$
in cases (ii) and (iii). Here a simple computation based on the known actions
of $H$ completes the proof of the Lemma. \qed
\medskip

Theorem \ref{Th:primitive} now follows by combining the above two lemmas.
In fact we obtain, for all primitive subgroups $H \ne A_n,S_n$,
\[
D_H \le 2 n^{-m(H)/84} \le 2 n^{-(\sqrt{n}-1)/168}.
\]

The remainder of this section is devoted to reducing Theorem \ref{thm:c} to
Theorem \ref{th:size}.

\begin{lem}\label{lem:sizeC}
Let $C$ be a conjugacy class in $S_n$ consisting of elements of support $k$.
Then $|C| \geq c \binom {n}{k} \sqrt{k!}\cdot k ^{-\frac{1}{2}}$, where
$c$ is an absolute positive constant.
\end{lem}
\paragraph{Proof:}
There a $\binom {n}{k}$ ways to chose the subset $S \subset \{1, \ldots,n\}$ of
letters moved by an element $h \in C$. Given the subset $S$, $h|_S$ is a fixed
point free permutation of degree $k$. The number of such permutations with a
given cycle structure is minimal in the case of a fixed point free involution
and is in this case equal to $k!/2^{\frac{k}{2}} (k/2)!$. Using Stirling's
formula, we see that this expression is at least $c \sqrt{k!}\cdot
k^{-\frac{1}{2}}$. Putting everything together the lemma follows. \qed

\begin{lem}\label{lem:last}
Let $H \le S_n$. Then
\[
D_H \leq  a \sum_{1 \le k \le n} |H_k| \binom {n}{k}^{-\frac{1}{2}}
(k!)^{-\frac{1}{4}}\cdot k^{\frac{1}{4}},
\]
where $a$ is some absolute constant.
\end{lem}

\paragraph{Proof:}
We use part 2 of Proposition \ref{Th:main}:
\[
D_H \le \sum_{1 \ne h \in H} |h^G|^{-1/2}.
\]
By Lemma \ref{lem:sizeC} we conclude that
\[
\sum_{h \in H_k} |h^G|^{-\frac{1}{2}} \leq c^{-1/2}
|H_k| \binom {n}{k}^{-\frac{1}{2}} (k!)^{-\frac{1}{4}}\cdot k^{\frac{1}{4}}.
\]
The result follows.
 \qed

Suppose now that Theorem \ref{th:size} holds and let $m= m(H)$.
Substituting
\[
|H_k|\leq n^{-\varepsilon m} \binom {n}{k}^{\frac{1}{2}}
(k!)^{\frac{1}{4}}
\]
in Lemma \ref{lem:last} we obtain
\[
D_H \leq a n^{-\varepsilon m} \sum_{1 \leq k \leq n}
k^{\frac{1}{4}}\leq a n^{-\varepsilon m} \cdot n^{\frac{5}{4}}.
\]

Therefore, if $m$ is unbounded, $D_H$ is smaller than any inverse
polynomial in $n$, and hence $H$ is indistinguishable. Moreover, assuming
$D_H \ge n^{-c}$ (and $n^{3/4} \ge a$ as we may) we obtain
$\varepsilon m - 2 \le c$, and
so
\[
m \le 2/\varepsilon + c/\varepsilon.
\]
Hence Theorem \ref{thm:c} follows from
Theorem \ref{th:size}.

\bigskip
\bigskip

\section{Bounds on the group size in terms of the minimal degree}\label{s:5}

In this section we prove Theorem \ref{thm:A}. It extends a theorem of Martin
Liebeck \cite{Li1,Li2} which bounds the order of transitive groups with large
minimal degree.

We call $H$ a \emph{subdirect product subgroup} of $S^t$ if it is a subdirect
product of $S_1 \times \dots \times S_t$ where all the $S_i$ are isomorphic
to~$S$. Such an $H$ is called a \emph{diagonal subgroup} if it is isomorphic
to~$S$.

\begin{lem}
\label{lem:1} Let $S$ be a non-abelian simple group and $H$ a subdirect product
subgroup of $S^t \cong S_1 \times \dots \times S_t$.

{\rm 1)} Then there is a partition of the set of indices $\{1, \dots, t\}$ and
for each part, say $\{i_{j_1},\dots, i_{j_k}\}$, a diagonal subgroup $D_j$ of
$S_{i_{j_1}} \times \dots \times S_{i_{j_k}}$ such that $H$ is a direct product
of the subgroups~$D_j$.

{\rm 2)} Assume that $S\cong \text{\rm Alt}(k)$ for some $k \geq 7$ and let $D$
be a diagonal subgroup of~$S^t$. Let $d = (d_1, \dots, d_t)$ be an element of
$D$ such that $d_1$ is a $3$-cycle. Then all the $d_i$ are $3$-cycles.
\end{lem}

\begin{proof}
1. This is a standard result.

2. This follows from the fact that the set of $3$-cycles is invariant under
automorphisms of $\text{\rm Alt}(k)$ if $k \geq 7$ \cite[Lemma 8.2.~A]{DM}.
\end{proof}

Let $H$ be a permutation group with minimal degree $m = m(H)$. Denote by
$\Omega_1, \dots, \Omega_r$ the orbits of $H$ and set $t = \max|\Omega_i|$. Let
${\mathcal B}_i = \{B_{i_1}, \dots, B_{i_{k_i}}\}$ a system of blocks of
imprimitivity for the action of $H$ on $\Omega_i$ such that $k_i > 1$ is
minimal (if $H$ acts on $\Omega_i$ as a primitive group, then $k_i =
|\Omega_i|$). Denote by $K_i$ the kernel of the action of $H$ on ${\mathcal
B}_i$ and the size of the blocks in ${\mathcal B}_i$ by $b_i$. Set ${\mathcal
B} = \bigcup\limits^r_{i = 1} {\mathcal B}_i$, $K = \bigcap\limits^r_{i = 1}
K_i$ and $x = \sum\limits^r_{i = 1} (k_i - 1)$. Note that $K$ has at least $r +
x$ orbits.

\begin{proposition}
\label{prop:2} $|H / K| \leq 5^x\,  t^{3n/m}$.
\end{proposition}

\begin{proof}
$H$ acts on ${\mathcal B}_i$ as a primitive permutation group $P_i \cong H /
K_i$ of degree $k_i$. If $P_i$ does not contain $\text{\rm Alt}(k_i)$, then, by
a result of Praeger and Saxl, \cite{PS} we have $|P_i| \leq 4^{k_i}$. Together
with some trivial computation for small values of~$k_i$ this implies $|P_i|
\leq 5^{k_i - 1}$.

Denote by $S$ the intersection of all the $K_i$ for which $|P_i| \leq 5^{k_i -
1}$ holds. Then $S$ acts on each ${\mathcal B}_i$ either as a trivial group or
as a group containing $\text{\rm Alt}(k_i)$ where $k_i \geq 7$. Without loss of
generality one can assume that $S$ acts trivially on ${\mathcal B}_i$ exactly
if $i > q$. The group $A = (S/K)'$ is a subdirect product subgroup of
$\text{\rm Alt}(k_1)\times \dots \times \text{\rm Alt}(k_q)$. Denoting by
$\overline A$ the inverse image of $A$ in $S$ we see that $|H / \overline A|
\leq 5^x$ holds.

To complete the proof it is enough to show that
\[
|\overline A / K| = |A| \leq t^{3n/m}.
\]

It follows from Lemma~\ref{lem:1} that $A$ is a direct product of diagonal
subgroups~$A_j$. Each $A_j$ acts as an alternating group $\text{\rm Alt}(n_j)$
on some systems of blocks ${\mathcal B}_i$ with $n_j = k_i$, trivially on the
rest and is isomorphic to $\text{\rm Alt}(n_j)$.

We claim that the sum of the block-sizes $b_i$ corresponding to $A_j$ is at
least $m/3$. To simplify notation we assume that $A_j$ acts trivially on
${\mathcal B}_i$ exactly if $i > p$. By Lemma~\ref{lem:1} there is an element
$a_j$ of $A_j$ which acts as a $3$-cycle on each ${\mathcal B}_i$ for $i \leq
p$. This element corresponds to an element $\overline{a_j}$ of $\overline A$
which moves at most $3 \sum\limits^p_{i = 1} b_i$ elements. Hence $3
\sum\limits^p_{i = 1} b_i \geq m$ as claimed.

It follows that each $A_j$ moves at least $n_j m / 3$ points. This implies that
the sum of the $n_j$ for all diagonal subgroups $A_j$ is at most $3n/m$. Each
$A_j$ has order $\frac12 n_j! \leq t^{n_j}$. Hence $|A| \leq t^{3n/m}$ as
required.
\end{proof}

We are now ready to prove Theorem \ref{thm:A}:

\begin{proof}[Proof of Theorem \ref{thm:A}]
Set $\ell = \min(m, \log_2 n)$. We have to show that $|G| \leq
n^{10\frac{n}{\ell}}$. Denote by $\Delta_1, \dots, \Delta_t$ the orbits of~$G$.
Let $\mathcal D_i = \{D_{i1}, \dots, D_{i h_i} \}$ be a system of blocks of
imprimitivity for the action of $G$ on $\Delta_i$, such that the block size
$d_i$ is at least $\ell$ and $d_i$ is as small as possible with this
restriction (if there are no proper blocks of size $\geq \ell$ then we set
$D_{i1} = \Delta_i$). $G$ acts on $\mathcal D = \bigcup\limits^t_{i = 1}
\mathcal D_i$ as a permutation group of degree at most $n / \ell$. Hence the
kernel $H$ of the action has index $\leq n^{\frac{n}{\ell}}$ in~$G$.

Denote by $\Omega_1, \dots, \Omega_r$ the orbits of $H$ and let $\mathcal B_1,
\dots, \mathcal B_r$ be systems of imprimitivity as in
Proposition~\ref{prop:2}. By the construction of $H$ it is clear that we have
$b_i < \ell$ for each~$i$. Applying Proposition~\ref{prop:2} we obtain a
subgroup $K$ of index $\leq 5^x n^{3n/m}$ such that $K$ has at least $r + x$
orbits and each orbit has size $< \ell$.

We apply Proposition~\ref{prop:2} to $K$ to obtain a subgroup $K_1$ of index
$\leq 5^{x_1} \cdot \ell^{3n/m}$ in $K$, which has at least $r + x + x_1$
orbits, each of size $\leq \frac{\ell}{2}$.

Continuing in this fashion we obtain a descending series of subgroups $K > K_1
> K_2 > \dots > K_v = 1$.
The maximal size of an orbit of $K_i$ is at most $\ell / 2^i$, hence the above
series of subgroups has length $v \leq \log_2 \ell$.

Since $K_i$ has at least $r + x + x_1 + \dots + x_i$ orbits we have $x + x_1 +
\dots + x_v \leq n$. Hence $|H| = |H / K| \cdot |K / K_1| \prod\limits^{v -
1}_{i = 1} |K_i / K_{i + 1} | \leq 5^n n^{3n/m} \cdot (\ell^{3n/m})^v \leq 5^n
\cdot n^{3n/\ell} \cdot 2^{3n\left( \frac{(\log \ell)^2}{\ell} \right)} \leq
5^n n^{3n/\ell} \cdot 2^{3n \cdot 9/8} \leq n^{3n/\ell} \cdot 2^{6n}$.
Therefore we have $|G| \leq n^{4n/\ell} \cdot 2^{6n} \leq n^{10\frac{n}{\ell}}$
as required.
\end{proof}

\bigskip
\bigskip

\section{Counting elements of given support}\label{s:6}

This section, which is the longest in this paper, is devoted to the proof of
Theorem \ref{th:size}. The main ingredients of the proof are Theorem
\ref{thm:A} and Proposition \ref{Th:primitive}.

We will use the following inequality many times.

\begin{proposition}
\label{prop:1} Let $x, y, n$ be positive integers such that $x + y \leq n$.
Then ${n \choose x}{n \choose y} \leq {n \choose x + y} 2^{2(x + y)}$ holds.
\end{proposition}

\begin{proof}
In fact we claim that the stronger inequality ${n \choose x} {n\choose y} \leq
{n \choose x + y} {x + y\choose y}^2$ holds. This is equivalent to
\[
\frac{n(n - 1)\dots (n - x + 1)n(n - 1) \dots (n - y + 1)} {n(n - 1) \dots (n -
x - y + 1)} \leq {x + y \choose y}
\]
which is equivalent to
\[
\frac{n(n - 1) \dots (n - y + 1)}{(n - x) \dots (n - x - y + 1)} \leq \frac{(x
+ y) (x + y - 1) \dots (x + 1)}{y!}
\]
But this follows by multiplying the inequalities
\[
\frac{n - t}{n - x - t} \leq \frac{x + y - t}{y - t} \quad \text{ for } \ t =
0,1,\dots, y - 1.
\]
These latter inequalities follow from $x + y \leq n$.
\end{proof}

To avoid some technical difficulties we first prove Theorem \ref{th:size}
directly in the case when $k$ is very large.

\begin{lem}
\label{lem:2} Let $H$ be a permutation group of degree $n$ and minimal degree
$m \geq 100\,000$. Assume that $k \geq n^{\frac23 + \frac1{100}}$ and $k \geq
2^{100\,000}$. Then there exists a constant $\varepsilon > 0$ such that $|H_k|
\leq {n\choose k}^{\frac12} (k!)^{\frac14} n^{-\varepsilon m}$ holds.
\end{lem}

\begin{proof}
We have to count elements $h \in H$ with $\text{\rm supp}(h) = k$. There are at
most ${n\choose k}$ choices for $\text{\rm supp}(h)$ and given this by
Theorem~\ref{thm:A} there are at most $k^{\frac{k}{10\,000}}$ choices for $h$
itself. We have to show that
\[
{n \choose k}k^{\frac{k}{10\,000}} \leq {n\choose k}^{\frac12}(k!)^{\frac14}
n^{-\varepsilon m}.
\]
This is equivalent to
\[
{n\choose k} k^{\frac{k}{5000}} \cdot n^{2\varepsilon m} \leq (k!)^{\frac12}
\]
which follows from
\[
n^k \cdot k^{\frac{k}{5000}} \cdot n^{2\varepsilon k} \leq (k!)^{\frac32}.
\]
This in turn is implied by
\[
n^k \cdot k^{\frac{k}{5000}} k^{3\varepsilon k} \leq
\left(\frac{k}{e}\right)^{\frac32 k}
\]
which reduces to
\[
n^{\frac23} \big(e \cdot k^{\frac1{7500} + 2\varepsilon} \big) \leq k
\]
which follows from our conditions if $\varepsilon$ is small enough.
\end{proof}

We now fix $a\geq 10\,000$ such that if $H$ is a primitive permutation group of
degree $n \geq a$ not containing $\text{\rm Alt}(n)$, then $m(H) \geq 100$ and
$|H_k| \leq n^{k/7}$. This is possible by \cite{Babai:81a} and Lemma
\ref{lem:oneseventh} above.

Next, we introduce some notation which will be used in the rest of this
section. Let $G$ be a permutation group of degree $n$ with no fixed points.
Denote by $\Omega_1, \Omega_2, \dots$, the orbits of~$G$. Let $\mathcal B_i =
\{B_{i1}, B_{i2}, \dots \}$ be a system of blocks of imprimitivity for the
action of $G$ on $\Omega_i$, such that $|B_{i1}| \geq 2$ is minimal. Then the
setwise stabiliser of the blocks $B_{ij}$ in $G$ acts as some primitive group
$P_{ij}$ on $B_{ij}$. The $P_{ij}$ are permutation equivalent for $i$ fixed.

We partition the set of blocks ${\mathcal B} = \bigcup {\mathcal B}_i$ into $3$
subsets as follows. Denote by ${\mathcal S} = \{ S_1, S_2, \dots\}$ the set of
blocks of size $< a$. Denote by $\mathcal A = \{ A_1, A_2, \dots \}$ the set of
blocks $B_{ij}$ in ${\mathcal B} \setminus {\mathcal S}$ for which $P_{ij}$
contains $\text{\rm Alt} (B_{ij})$, and denote by $\mathcal L = \{ L_1, L_2,
\dots, \}$ the set of the remaining blocks. Set $S = \bigcup S_i$, $A = \bigcup
A_i$ and $L = \bigcup L_i$. It is clear that any $g \in G$ fixes the sets $S$,
$L$ and~$A$. We denote the action of $g \in G$ on a set $X$ (fixed by~$g$) by
$g_X$ and the action of $G$ on a fixed set $X$ by $G_X$.

Our next lemma shows that in a sense there are not too many possibilities for
the action of some $g \in G$ on the set $S\cup L$.

\begin{lem}
\label{lem:4} {\rm 1)} The number of pairs $(\text{\rm supp} (g_S), g_L)$ for
permutations $g$ with $|\text{\rm supp}(g)| = k$, $|\text{\rm supp}(g_L)| = x$
and $|\text{\rm supp} (g_A)| = y$ is at most $\left({n \atop \left[\frac{k - x
- y}{2} \right]}\right)2^{ak} \cdot n^{x \left(\frac17 + \frac1{100}\right)}$.

{\rm 2)} Given $\text{\rm supp}(g_S)$, the number of possible actions $g_S$ is
at most $a^{k - x - y} \left[ \frac{k - x - y}{2}\right]!$. In fact this is an
upper bound for the number of possible actions on $\text{\rm supp}(g_S)$ of
elements $h$ which fix $\text{\rm supp}(g_S)$.
\end{lem}

\begin{proof}
If $g$ moves a point of some block, then it moves at least two points of the
block. Hence the number $t$ of blocks in $\mathcal S$ which contain points from
$\text{\rm supp}(g)$ is at most $\left[\frac{k - x - y}{2}\right]$. These
blocks can be chosen in at most ${n\choose t}$ ways. Given these blocks the
number of choices for $\text{\rm supp} (g_S)$ is at most $(2^a)^t$.

Note that $g_S$ (or $h \in G$ fixing $\text{\rm supp}(g_S)$) moves $a_1, a_2,
\dots, a_t$ given points of the chosen blocks in at most $a_1! a_2! \dots a_t!
\cdot t! \leq \prod\limits^t_{i = 1} a^{a_i} \cdot t! \leq a^{k - x - y} \left[
\frac{k - x - y}{2} \right]!$ ways, proving~2).

Each block in $\mathcal L$ which contains points of $\text{\rm supp}(g)$
contains at least 100 such points (by the choice of $a$, see the notation
introduced after Lemma~\ref{lem:2}), hence the number $\ell$ of such blocks is
at most $x/100$. These blocks can be chosen in at most ${n \choose \ell} \leq
n^{\frac{x}{100}} / \ell!$ ways.

There are $\ell_1 \leq \ell$ blocks from $\mathcal L$ fully contained in
$\text{\rm supp}(g)$ and these can be chosen in at most $2^\ell$ ways.

By our assumption on the blocks in $\mathcal L$ and the Praeger--Saxl theorem
\cite{PS} the stabilisers of a block $B_{ij}$ in $\mathcal L$ can act on the
block in at most $4^{|B_{ij}|}$ ways. This implies that the stabiliser of the
union of the above blocks can act on this union in at most $4^x \ell_1!$ ways.
Hence this is an upper bound for the number of actions of $g$ on the blocks
contained in $\text{\rm supp}(g)$.

Assume that on the remaining blocks (which are as sets fixed by~$g$) $g$ acts
as a permutation of degree $x_1, x_2, \dots$. The number $x_1, x_2, \dots$ can
be chosen in at most $2^x$ ways. Given these numbers the number of actions of
$g$ on these remaining blocks can be chosen in at most $n^{x_1/7} \cdot
n^{x_2/7} \cdots \leq n^{x/7}$ ways by Lemma \ref{lem:oneseventh}.

Altogether the number of choices for $\text{\rm supp}(g_S)$ and $g_L$ is at
most
\[
{n \choose t} 2^{at} \big(n^{\frac{x}{100}} / \ell! \big) 2^\ell 4^x \ell_1!
2^x \cdot n^{x/7} \leq \left( {n \atop \left[ \frac{k - x - y}{2}
\right]}\right) n^{\frac{x}{7} + \frac{x}{100}} 2^{ak}
\]
as required.
\end{proof}

\begin{corollary}
\label{cor:5} The number of pairs $(\text{\rm supp} (g_S), g_L)$ for
permutations $g$ with $|\text{\rm supp}(g)| = k$ and $|\text{\rm supp} (g_A)| =
y$ is at most
\[
{n \choose k}^{\frac12} \left[\frac{y}2\right]! n^{-\frac{y}{2}} \cdot 2^{(a +
4)k} \quad \text{if } k \leq n^{\frac23 + \frac{1}{100}}
\]
and $n$ is sufficiently large.
\end{corollary}

\begin{proof}
We first claim that the number of permutations $g$ considered is at most
$\frac1n\left({n\atop \left[\frac{k - y}{2}\right]} \right) 2^{(a + 1)k}$. By
Lemma~\ref{lem:4} it is sufficient to prove that for all $x \leq k$ we have
\[
\left({n\atop \left[\frac{k - x - y}{2}\right]}\right) 2^{ak} n^{x\left(
\frac17 + \frac1{100}\right)} \leq \frac1{kn} \left( {n \atop \left[\frac{k -
y}{2}\right]}\right)2^{(a + 1)k}.
\]

This is obvious if $x = 0$, otherwise we have $x \geq 100$. By
Proposition~\ref{prop:1}
\[
\left({n \atop \left[\frac{k - x - y}{2} \right]}\right) \left({n \atop
\left[\frac{x}{2}\right]}\right) \leq \left( {n \atop \left[ \frac{k -
y}{2}\right]} \right) 2^k
\]
holds, hence it is enough to show that $n^{x\left(\frac17 + \frac1{100}\right)
+ 2} \leq \left({n \atop \left[ \frac{x}{2}\right]} \right)$. But this follows
using $100 \leq x \leq k \leq n^{\frac23 + \frac1{100}}$.

Using Proposition~\ref{prop:1} we obtain that

\begin{align*}
\frac1n \left({n \atop \left[\frac{k - y}{2}\right]}\right) 2^{(a + 1)k} &\leq
\frac1{n} \left({n \atop \left[ \frac{k}{2} \right]} \right) \left( {n\atop
\left[\frac{y}{2}\right]} \right)^{-1}
2^{(a + 2)k} \\
&\leq \frac1n {n \choose k}^{\frac12} \left({n \atop \left[ \frac{y}{2}\right]}
\right)^{-1} 2^{(a + 3)k} \leq \frac1n {n\choose k}^{\frac12}
\left[\frac{y}{2}\right]!
n^{-\left[\frac{y}{2}\right]} 2^{(a + 4)k} \\
&\leq {n\choose k}^{\frac12} \left[\frac{y}{2}\right]! n^{-\frac{y}{2}} 2^{(a +
4)k}
\end{align*}
proving the corollary.
\end{proof}

The most difficult part of the proof of Theorem~\ref{th:size} is when $y$ is
large compared to~$m$. The following result implies Theorem~\ref{th:size} in
the case when this holds and moreover $k!$ is large compared to~$n^y$.

\begin{lem}
\label{lem:6} Assume that $m \geq 100\,000$, $k \leq n^{\frac23 +
\frac1{100}}$, $n^{3y} \leq k!$ and $k$ is sufficiently large (in particular $k
\geq 2^{100\,000}$). Then the number of permutations $g$ with $|\text{\rm supp}
(g)| = k$ and $|\text{\rm supp}(g_A)| = y$ is at most ${n \choose k}^{\frac12}
(k!)^{\frac14} n^{-\frac{y}{60}}$.
\end{lem}

\begin{proof}
The number of choices for $\text{\rm supp}(g_A)$ is at most ${n \choose y}$.
Hence by Corollary~\ref{cor:5} the number of choices for $\text{\rm supp}(g)$
is at most
\[
{n\choose k}^{\frac12} \left[\frac{y}{2}\right]! n^{-\frac{y}{2}} 2^{(a + 4)k}
{n \choose y} \leq {n \choose k}^{\frac12} n^{\frac{y}{2}} 2^{(a + 4)k}.
\]

Using Theorem~\ref{thm:A} we see that the number of choices for $g$ is at most
\[
{n\choose k}^{\frac12} n^{\frac{y}{2}} 2^{(a + 4)k} k^{\frac{k}{10\,000}} \leq
{n\choose k}^{\frac12} (k!)^{\frac15} \cdot k^{\frac{k}{10\,000}} \cdot 2^{(a +
4)k} \cdot n^{-\frac{y}{60}}.
\]

If $k$ is large enough (compared to the constant $a$) then $(k!)^{\frac{1}{20}}
\geq k^{\frac{k}{10\,000}} \cdot 2^{(a + 4)k}$ and our statement holds.
\end{proof}

Next we describe an important subgroup of~$G$. Consider the set consisting of
the points in $S$ and $L$ and the blocks in~$\mathcal A$. Let $K$ be the kernel
of the action of $G$ on this set. By definition $K$ fixes all the points
outside~$A$. Moreover, if $A_i \in \mathcal A$, then the action $K_i$ of $K$ on
$A_i$ is a normal subgroup of the action of the stabiliser of $A_i$ in $G$,
hence it is either $\text{\rm Sym}(A_i)$, $\text{\rm Alt}(A_i)$ or $1$.

Without loss of generality one can assume that $K$ acts trivially on $A_i$
exactly if $i > q$. Now $K$ is a subdirect product of the $K_i$, therefore its
commutator subgroup $K'$ is a subdirect product subgroup of $\text{\rm Alt}\,
(A_1) \times \dots \times \text{\rm Alt}\, (A_q)$. Hence by Lemma~\ref{lem:1}
$K'$ is a direct product of diagonal subgroups $D_j$. Each $D_j$ acts as an
alternating group $\text{\rm Alt}\, (n_j)$ on some blocks $A_i$ of size~$n_j$.
By Lemma~\ref{lem:1} $D_j$ contains an element $d_j$ which acts as a 3-cycle on
each of the corresponding~$A_i$. Hence $D_j$ acts non-trivially on at least
$\frac{m}{3}$ blocks $A_i$ (since $|\text{\rm supp} (d_j)| \geq m$). Now $K$ is
a subgroup of the normaliser $N$ of $K'$ in $\prod\limits^q_{i = 1} \text{\rm
Sym}(A_i)$. Clearly $N$ is a direct product of groups $N_j \geq D_j$ where
$N_j$ is isomorphic to $\text{\rm Sym}(n_j)$ and contains $D_j \cong \text{\rm
Alt}(n_j)$ in a natural way.

\begin{proposition}
\label{prop:7} There are at most $n^{\frac{3h}{m}}$ elements $g$ of $K$ with
$|\text{\rm supp}(g)| = h$ (where $m = m(G)$).
\end{proposition}

\begin{proof}
We have a unique decomposition $g = g_1 g_2 \dots$ where $g_j \in N_j$. Let us
choose for each $j$ a block on which $N_j$ acts non-trivially. It is clear that
$g_j$ is determined uniquely by its action on the chosen block. Therefore $g$
is determined by its action on the union $U$ of the chosen blocks.

It follows by the above discussion that $|\text{\rm supp}(g) \cap U| \leq
\frac{3h}{m}$. Hence the number of choices for $g$ is at most
$|U|^{\frac{3h}{m}} \leq n^{\frac{3h}{m}}$.
\end{proof}

\begin{proposition}
\label{prop:8} Assume that $m \geq 100\,000$. Then the number of permutations
$g$ with $g_{S \cup L}$ fixed and $\text{\rm supp} (g_A) = y$  is at most
$n^{y/5000}$.
\end{proposition}

\begin{proof}
The coset $gK$ is determined by $g_{S\cup L}$ and the action of $g$ on the
blocks in~$\mathcal A$. Now $g$ can move at most $t \leq \frac{y}{a}$ blocks in
$\mathcal A$.

The number of choices for these blocks is less than ${n/a\choose t}$ and given
these blocks the number of ways $g$ can act on them is at most $t!$. Hence $g$
can act in at most $\left(\frac{n}{a}\right)^{\left[\frac{y}{a}\right]} +
\left(\frac{n}{a}\right)^{\left[ \frac{y}{a}\right] - 1} + \dots \leq
n^{\frac{y}{a}}$ ways on $\mathcal A$. If $gK$ contains another element $f$
with $|\text{\rm supp}(f)| = k$ and $|\text{\rm supp}(f_A)| = y$, then $gf^{-1}
\in K$ and $|(\text{\rm supp}(gf^{-1})| \leq 2y$. Hence by
Proposition~\ref{prop:7} there are at most $n^{\frac{6y}{m}} \leq
n^{\frac{y}{10\,000}}$ such elements $gf^{-1}$. Of course $g$ and $gf^{-1}$
determines~$f$. Altogether we see that the number of elements $g$ considered is
at most $n^{\frac{y}{5000}}$.
\end{proof}

\begin{rema}
As the proof shows (see also the proof of Proposition~\ref{prop:7} and the
preceding discussion) the conclusion of Proposition~\ref{prop:8} holds under
the much weaker assumption that all elements of order $3$ in $G$ move at least
$100\,000$ points.
\end{rema}

\begin{proposition}
\label{prop:9} Assume that $m \geq 100\,000$, $k \leq n^{\frac23 +
\frac1{100}}$, $y \neq 0$ and $n$ is sufficiently large. Then the number of
permutations $g \in G$ with $|\text{\rm supp} (g)| = k$ and $|\text{\rm
supp}(g_A)| = y$ is at most ${n \choose k}^{\frac12} n^{-\frac{y}{2} +
\frac{y}{5000}} k! 2^{(a + 4)k}$.
\end{proposition}

\begin{proof}
By Corollary~\ref{cor:5} the number of possibilities for $\text{\rm supp}
(g_{S\cup L})$ is at most\break ${n \choose k}^{\frac12}
\left[\frac{y}{2}\right]! n^{-\frac{y}{2}} \cdot 2^{(a + 4)k}$. Therefore the
number of possibilities for $g_{S \cup L}$ is at most
\[
{n \choose k}^{\frac12} n^{-\frac{y}{2}} 2^{(a + 4)k} \left[\frac{y}{2}\right]!
(k - y)! \leq {n \choose k}^{\frac12} n^{-\frac{y}{2}} 2^{(a + 4)k} \cdot k!.
\]
Hence by Proposition~\ref{prop:8} the number of choices for $g$ is at most
\[
{n \choose k}^{\frac12} n^{-\frac{y}{2} + \frac{y}{5000}} \cdot k! 2^{(a + 4)k}
\]
as required.
\end{proof}

The next result as a counterpart of Lemma~\ref{lem:6} deals with the case when
$n^y$ is large compared to $k!$ (and $y$ is large compared to~$m$).

\begin{corollary}
\label{cor:10} Assume that $k \leq n^{\frac23 + \frac1{100}}$, $n^{\frac{y}{8}}
\geq k!$ and $m$ is sufficiently large. Then the number of permutations $g$
with $|\text{\rm supp}(g)| = k$ and $\text{\rm supp}(g_A)| = y$ is at most ${n
\choose k}^{\frac12} n^{-\frac{y}{5}}$.
\end{corollary}

\begin{proof} We have $m \leq k \leq n$, hence if $m$ is large enough
Proposition~\ref{prop:9} is applicable. Moreover, we have $2^{(a + 4)k} \leq
k!$ if $m$ is large enough (compared to the fixed constant~$a$).

Hence in this case we have
\[
{n\choose k}^{\frac12}n^{-\frac{y}{2} + \frac{y}{5000}} \bigl(k! 2^{(a +
4)k}\bigr) \leq {n \choose k}^{\frac12} n^{-\frac{y}{2} + \frac{y}{5000} +
\frac{y}{4}} \leq {n \choose k}^{\frac12} n^{-\frac{y}{5}}.
\]
 \end{proof}
To deal with the case when $k!$ and $n^y$ are ``almost equal'' we have to
introduce further ideas and notation. We call a pair of the form $(\text{\rm
supp}(g_S), g_L)$ \emph{thick} if the elements $g$ which correspond to it act
in at least $(k!)^{\frac16}$ different ways on $\text{\rm supp}(g_S)$ and call
a pair \emph{thin} otherwise.

\begin{proposition} \label{prop:11}
Assume that $m \geq 100\,000$, $2^{200 a} \leq k \leq n^{\frac23 +
\frac1{100}}$, $y \neq 0$ and $n$ is sufficiently large. Then the number of
permutations $g$ with $|\text{\rm supp}(g)| = k$ and $|\text{\rm supp} (g_A)| =
y$ for which $(\text{\rm supp}(g_S), g_L)$ is thin is at most ${n \choose
k}^{\frac12} (k!)^{\frac16 + \frac1{100}}$.
\end{proposition}

\begin{proof} By Corollary~\ref{cor:5} the number of possibilities for the pair
$(\text{\rm supp}(g_S), g_L)$ is at most ${n \choose k}^{\frac12}
\left[\frac{y}{2}\right]! n^{-\frac{y}{2}} \cdot 2^{(a + 4)k}$. Hence the
number of possibilities for $g_{S\cup L}$ is at most
\[
 {n\choose k}^{\frac12} \left[\frac{y}{2}\right]! n^{-\frac{y}{2}} \cdot 2^{(a + 4)k} (k!)^{\frac16}
\leq {n \choose k}^{\frac12} \left[\frac{y}{2} \right]!
n^{-\frac{y}{2}}(k!)^{\frac16 + \frac1{100}}
\]
(we used the condition $200 a \leq \log k$). Using Proposition~\ref{prop:8} we
see that the total number of elements $g$ considered is at most
\[
 {n\choose k}^{\frac12}(k!)^{\frac16 + \frac1{100}} n^{-\frac{y}{2} + \frac{y}{5000}}
\left[ \frac{y}{2} \right]! \leq {n\choose k}^{\frac12} (k!)^{\frac16 +
\frac1{100}}
\]
(using $y \leq k \leq n^{\frac23 + \frac1{100}}$).
 \end{proof}
 \begin{proposition} \label{prop:12} Let $(\text{\rm supp}(g_S), g_L)$ be a thick pair.
Denote the action of (the stabiliser of $\text{\rm supp}(g_S)$ in) $G$ on
$\text{\rm supp}(g_S)$ by $H$. There is an element $\gamma$ which corresponds
to this pair such that the centraliser of $\gamma_S$ in $H$ has order at most
\[ (5a)^{k - x - y} \left[ \frac{k - x - y}{2} \right]! \big/ (k!)^{\frac16}.
\]
\end{proposition}

 \begin{proof} By Lemma~\ref{lem:4}(2) $H$ has order at most $a^{k - x - y}
\left[ \frac{k - x - y}{2} \right]!$. By a result of Kov\'acs and Robinson
\cite{KR} the number $k(H)$ of conjugacy classes of the permutation group $H$
is at most $5^{k - x - y}$. Using a well-known identity we obtain
\[ \sum_{h\in H} C_H(h) = k(H) |H| \leq (5a)^{k - x - y} \left[ \frac{k - x - y}{2}\right]!
\, .
\]
Since by definition we have at least $(k!)^{\frac16}$ choices for $g_S \in H$,
at least one of them has small centraliser as required.
\end{proof}

\begin{proposition} \label{prop:13}
Assume that $m \geq 100\,000$, $k \geq 2^{100\,000}$ and $y \neq 0$. Let\break
$(\text{\rm supp}(g_S), g_L)$ be a thick pair and $\gamma$ a corresponding
permutation with small centraliser as above. The number of elements $g$ which
correspond to this pair and satisfy the condition
\[ |\text{\rm supp}(g_A) \cap \text{\rm supp}(\gamma_A)| \geq\frac{y}{100}
\]
is at most $a^k k^{\frac{k}{4} + \frac{k}{10\,000}} n^{0.4951 y} \big/ \left[
\frac{y}{2}\right]!$.
\end{proposition}

\begin{proof} The number of choices for the set $\text{\rm
supp}(g_A) \cap \text{\rm supp}(\gamma_A)$ is less than $2^y$. The number of
choices for the rest of $\text{\rm supp}(g_A)$ is at most ${n \choose [0.99\,
y]}$. Given these sets (and hence $\text{\rm supp}(g)$) by Theorem~\ref{thm:A}
the number of choices for $g$ is at most $k^{\frac{k}{10\,000}}$. It follows
that the number of choices for $g$ is less than
\[
2^k n^{0.99 y} k^{\frac{k}{10\,000}} \big/ \left[\frac{y}{2}\right]!\, .
\]
Another estimate for the number of possible choices for $g$ is the following.
The number of choices for $g_{S\cup L}$ is at most $a^{k - x - y} \left[\frac{k
- x - y}{2}\right]!$ by Lemma~\ref{lem:4}(2). Hence by Proposition~\ref{prop:8}
the number of choices for $g$ is less than
\[
a^{k - x - y} \left[ \frac{k - x - y}{2} \right]! n^{\frac{y}{5000}} \leq a^k
k^{\frac{k}{2}} n^{\frac{y}{5000}} \Bigm/ \left[\frac{y}{2}\right]!\, .
\]
A third estimate follows immediately from these; the number of choices for $g$
is at most
\[
 \Bigl(a^k \cdot k^{\frac{k}{2}} n^{\frac{y}{5000}} \cdot 2^k
\cdot n^{0.99 y} \cdot k^{\frac{k}{10\,000}}\Bigr)^{\frac12} \Bigm/
\left[\frac{y}{2}\right]! \leq a^k k^{\frac{k}{4} + \frac{k}{10\,000}}
n^{0.4951 y} \Bigm/ \left[ \frac{y}{2} \right]!
\]
as required.
\end{proof}

\begin{proposition}
\label{prop:14} Assume that $m \geq 100\,000$ and $k \geq 2^{100\,000}$. Let
$(\text{\rm supp}(g_S), g_L)$ be a thick pair and $\gamma$ a corresponding
permutation with small centralizer (as in Proposition~\ref{prop:12}). The
number of elements $g$ which correspond to this pair and satisfy
\[
|\text{\rm supp}(g_A) \cap \text{\rm supp}(\gamma_A)| \leq \frac{y}{100}
\]
is at most
\[ n^{\frac{y}{30}} \cdot k^{\frac{k}{3} + \frac{k}{10\,000}} (5a)^k \Bigm/
\left[\frac{y}{2} \right]!\, .
\]
 \end{proposition}

 \begin{proof} Let us consider the commutator $[\gamma, g]$. By \cite[Exercise
1.6.7]{DM} we have
\[
|\text{\rm supp}([\gamma, g]) \cap A| \leq 3 |\text{\rm supp}(g_A) \cap
\text{\rm supp}(\gamma_A)| \leq \frac{3y}{100}.
\]
Hence the number of choices for $\text{\rm supp}([\gamma, g]) \cap A$ is at
most $n^{\frac{3y}{100}}$. Note that $\text{\rm supp}([\gamma, g]) \cap (S \cup
L) \leq \text{\rm supp} (\gamma_{S \cup L})$ (which is fixed). Using
Theorem~\ref{thm:A} we obtain that the number of choices for $[\gamma, g]$ is
at most $n^{\frac{3y}{100}} \cdot k^{\frac{k}{10\,000}}$. This commutator,
together with $\gamma$, determines $g^{-1} \gamma g = \gamma[\gamma, g]$. If
$h$ is another element with $h^{-1} \gamma h = g^{-1} \gamma g$, then $gh^{-1}$
centralises $\gamma$. Hence by the choice of $\gamma$ in
Proposition~\ref{prop:12}) the number of possibilities for $h_S$ is less than
\[ (5 a)^k \left[\frac{k - y}{2}\right]! \Bigm/ (k!)^{\frac16}. \] Hence we
have at most $n^{\frac{3y}{100}} \cdot k^{\frac{k}{10\,000}} (5a)^k
\left[\frac{k - y}{2}\right]! \Bigm/ (k!)^{\frac16}$ choices for $g_{S \cup L}$
and given this, the number of choices for $g$ is at most $n^{\frac{y}{5000}}$
by Proposition~\ref{prop:8}. Therefore the number of choices for $g$ is at most
\[ n^{\frac{3y}{100}} \cdot n^{\frac{y}{5000}} (5a)^k k^{\frac{k}{10\,000}}
\left[\frac{k}{2}\right]! \Bigm/ (k!)^{\frac16} \left[ \frac{y}{2} \right]!
\leq n^{\frac{y}{30}} \cdot k^{\frac{k}{3} + \frac{k}{10\,000}} (5a)^k \Bigm/
\left[ \frac{y}{2}\right]! \]
\end{proof}

Our next result which builds on most of the earlier ones in this section
implies Theorem~\ref{th:size} if $y$ is large compared to~$m$.

\begin{lem}[Main Lemma] \label{lem:15}
Assume that $k \leq n^{\frac23 + \frac1{100}}$, $y \neq 0$ and $m$ is
sufficiently large. Then the number of permutations $g$ with $|\text{\rm
supp}(g)| = k$ and $|\text{\rm supp}(g_A)| = y$ is at most ${n \choose
k}^{\frac12} (k!)^{\frac14} n^{-\frac{y}{200}}$.
\end{lem}

\begin{proof} By Lemma~\ref{lem:6} and Corollary~\ref{cor:10} we may
assume that $n^{3y} \geq k! \geq n^{\frac{y}{8}}$. By Proposition~\ref{prop:11}
the number of permutations $g$ with a thin pair $(\text{\rm supp}(g_S), g_L)$
is at most \[ {n \choose k}^{\frac12}(k!)^{\frac16 + \frac1{100}} \leq {n
\choose k}^{\frac12} (k!)^{\frac14} (k!)^{-\frac1{20}} \leq {n \choose
k}^{\frac12} (k!)^{\frac14} n^{-\frac{y}{160}}. \] It remains to bound the
number of permutations $g$ with a thick pair. By Corollary~\ref{cor:5} the
number of possibilities for $(\text{\rm supp}(g_S), g_L)$ is at most ${n
\choose k}^{\frac12} \left[\frac{y}{2}\right]! n^{-\frac{y}{2}} \cdot 2^{(a +
4)k}$. Given this, by Propositions~\ref{prop:13} and \ref{prop:14} the number
of choices for $g$ is at most
\begin{align*} &\bigl(a^k k^{\frac{k}{4} + \frac{k}{10\,000}} n^{0.4951 y}
+ (5a)^k k^{\frac{k}{3} + \frac{k}{10\,000}} n^{\frac{y}{30}}\big) \Bigm/
\left[\frac{y}{2}\right]! \\
&\leq (10 a)^k (k!)^{\frac14} n^{\frac{3y}{10\,000}} \bigl(n^{0.4951 y} +
n^{\frac{y}{4}} \cdot n^{\frac{y}{30}}\bigr) \Bigm/
\left[\frac{y}{2} \right]! \\
&\leq (10 a)^k (k!)^{\frac14} n^{0.4954 y} \Bigm/ \left[\frac{y}{2} \right]!
\end{align*}
(we used the inequality $n^{3y} \geq \left(\frac{k}{e}\right)^k$). Hence the
total number of permutations $g$ with a thick pair is at most\break ${n \choose
k}^{\frac12} (k!)^{\frac14} n^{-0.006 y}\bigl((10 a)^k 2^{(a + 4)k}\bigr)$. If
$m$ and hence $k$ is large enough, then \[ (10 a)^k 2^{(a + 4)k} \leq
\frac{1}{2}(k!)^{\frac1{3000}} \leq \frac{1}{2}n^{\frac{y}{1000}}. \] Our
statement follows.
\end{proof}

Next we prove Theorem~\ref{th:size} in the case when $x$ is large compared
to~$m$.

\begin{proposition} \label{prop:16} Assume that $x \neq 0$, $n^{\frac23 + \frac1{100}}
\geq k \geq 2^{100\,000}$ and $m$ is sufficiently large. Then the number of
permutations $g$ with $|\text{\rm supp}(g)| = k$, $|\text{\rm supp}(g_L)| = x$
and $|\text{\rm supp}(g_A)| = y$ is at most ${n\choose k}^{\frac12}
(k!)^{\frac14} n^{-\frac{x}{20\,000}}$.
\end{proposition}

\begin{proof} If $y \geq \frac{x}{100}$, then our statement follows from the Main
Lemma. Assume now that $y \leq \frac{x}{100}$. By Lemma~\ref{lem:4} the number
of choices for $\text{\rm supp}(g)$ is at most $\left({n \atop \left[ \frac{k -
x - y}{2}\right]} \right) 2^{ak} n^{x\left(\frac17 + \frac1{100}\right)} \cdot
{n \choose y}$. Hence, by Theorem~\ref{thm:A} the number of choices for $g$ is
at most $\left({n \atop \left[\frac{k - x - y}{2} \right]} \right)2^{ak}
n^{x\left(\frac17 + \frac1{100}\right)} {n \choose y} k^{\frac{k}{10\,000}}$
(since we can assume that $m \geq 100\,000$). Using Proposition~\ref{prop:1}
and $y \leq \frac{x}{100}$ we see that this is at most
\begin{align*}
 &{n \choose k}^{\frac12} 2^{(a + 2)k} k^{\frac{k}{10\,000}} n^{x\left(\frac17 +
\frac1{100}\right)} {n \choose y} \Big/ \left( {n \atop \left[ \frac{x + y}{2}
\right]} \right)\\ &\leq {n \choose k}^{\frac12} 2^{(a + 2)k}
k^{\frac{k}{10\,000}} n^{x\left(\frac17 + \frac2{100}\right)} \Bigm/ \left({n
\atop \left[ \frac{x}{2}\right]}\right).
\end{align*}
If $m$ and hence $k$ is large enough compared to $a$, we have $2^{(a + 2)k}
k^{\frac{k}{10\,000}} \leq (k!)^{\frac14}$. Using $100 \leq x \leq k \leq
n^{\frac23 + \frac1{100}}$ we see that $n^{x\left(\frac17 + \frac2{100}\right)}
\Big/ \left({n \atop \left[\frac{x}{2} \right]} \right) \leq
n^{-\frac{x}{100}}$. Our statement follows.
\end{proof}

Let us return to the notation introduced after Lemma~\ref{lem:2}. If $S_i \in
{\mathcal S}$ is a small block, such that $g$ moves at least $3$ points of
$S_i$, then we denote $|\text{\rm supp}(g) \cap S_i|$ by $z_i$. We set $z(g) =
\sum z_i$ (for all such~$i$).

\begin{proposition} \label{prop:17}
Assume that $z \neq 0$, $n^{\frac23 + \frac1{100}} \geq k \geq 2^{100\,000}$
and $m$ is sufficiently large. Then the number of permutations $g$ with $z(g) =
z$ is at most ${n \choose k}^{\frac12}(k!)^{\frac14} n^{-\frac{z}{800\,000}}$.
\end{proposition}

\begin{proof} If $x \geq \frac{z}{40}$ or $y \geq \frac{z}{80}$, then our
statement follows from Lemma~\ref{lem:15} and Proposition~\ref{prop:16}. Assume
otherwise. If $g$ moves a point of some block, then it moves at least two
points of the block. Hence the number of blocks in ${\mathcal S}$ which contain
two points from $\text{\rm supp}(g)$ is at most $\left[\frac{k - z}{2}\right]$.
These blocks can be chosen in at most $\left({n \atop \left[\frac{k -
z}{2}\right]}\right)$ ways. The blocks in ${\mathcal S}$ which contain at least
$3$ points from $\text{\rm supp}(g)$ can be chosen in at most $\left({n \atop
\left[\frac{z}{3} \right]} \right)$ ways. Given these blocks the number of
choices for $\text{\rm supp}(g_S)$ is at most
\[
 \left({n \atop \left[\frac{k -z}{2} \right]}\right)
 \left({n \atop \left[\frac{z}{3} \right]}\right) 2^{az} a^{2 [\frac{k-z}{2}]}
\leq {n \choose k}^{\frac12} 2^{2k} \cdot 2^{ak} \left({n
\atop\left[\frac{z}{3}\right]}\right) \Bigm/ \left({n\atop \left\lceil
\frac{z}{2}\right\rceil}\right).
\]
Using $n^{\frac23 + \frac1{100}} \geq k \geq z$ we see that $\left({n \atop
\left[\frac{z}{3}\right]}\right) n^{\frac{z}{20}} \leq \left({n \atop
\left\lceil\frac{z}{2}\right\rceil}\right) 2^z$. Hence the number of choices
for $\text{\rm supp}(g)$ is at most
\[ {n\choose k}^{\frac12}  2^{(a + 3)k} n^{-\frac{z}{20}} n^{x + y}
\leq {n\choose k}^{\frac12} 2^{(a + 3)k} n^{-\frac{z}{80}}.
\]
The number of choices for $g$ itself is at most ${n\choose k}^{\frac12}
n^{-\frac{z}{80}} 2^{(a + 3)k} k^{\frac{k}{10\,000}}$ which is less than ${n
\choose k}^{\frac12} n^{-\frac{z}{80}}(k!)^{\frac14}$ if $k$ is large enough.
\end{proof}

Denote the number of small blocks $S_i \in {\mathcal S}$ fixed by $g$ such that
$|\text{\rm supp}(g) \cap S_i| = 2$ by $v(g)$. On these blocks $g$ acts as a
transposition.

\begin{proposition} \label{prop:18}
Assume that $n^{\frac23 + \frac1{100}} \geq k \geq 2^{100\,000}$ and $m$ is
sufficiently large, then the number of permutations $g$ with $v(g) = v \geq
\frac{m}{10}$ is at most ${n\choose k}^{\frac12} (k!)^{\frac14}
n^{-\frac{m}{800\,000\,000}}$.
\end{proposition}

\begin{proof} If $x + y + z \geq \frac{m}{1000}$, then our statement
follows from the previous results. Assume otherwise. Suppose first that $k!
\geq n^{\frac{m}{100}}$. The number of choices for small blocks $S_i$ with
$|\text{\rm supp}(g) \cap S_i| = 2$ is at most $\left({n \atop
\left[\frac{k}{2}\right]} \right)$. Hence the number of choices for all the
pairs $\text{\rm supp}(g) \cap S_i$ in these blocks is at most $\left({n \atop
\left[\frac{k}{2}\right]} \right) (a^2)^{\left[ \frac{k}{2}\right]} \leq {n
\choose k}^{\frac12} (2a)^k$. The number of choices for $\text{\rm supp}(g)$ is
then at most \[ {n\choose k}^{\frac12} (2a)^k n^{x + y + z} \leq {n\choose
k}^{\frac12} (2a)^k (k!)^{\frac1{10}}. \] Hence by Theorem~\ref{thm:A} the
number of choices for $g$ itself is at most ${n\choose k}^{\frac12} (2a)^k
(k!)^{\frac1{10}} k^{\frac{k}{10\,000}}$ which is less then ${n \choose
k}^{\frac12} (k!)^{\frac18}$ if $m$ is large enough. Therefore in this case the
number of permutations $g$ is at most ${n \choose k}^{\frac12} (k!)^{\frac14}
n^{-\frac{m}{800}}$. Suppose now that $k! \leq n^{\frac{m}{100}}$. To any
permutation $g \in G$ we assign a permutation $\overline g$ obtained by
``forgetting about'' $\left[\frac{m}{10}\right]$ transpositions in the small
blocks $S_j$ of the smallest index $j$ (which $g$ fixes and for which
$|\text{\rm supp}(g) \cap S_j| = 2$). Note that if $\overline g = \overline h$
then $|\text{\rm supp}(gh^{-1})| \leq \frac{m}{2}$, hence we have $g = h$. That
is $\overline g$ uniquely determines~$g$. The number of choices for $\text{\rm
supp}(\overline g)$ is at most \[ \left({n\atop \left[\frac{k}{2}\right] -
\left[ \frac{m}{10}\right]}\right) a^k n^{x + y + z} \leq {n\choose
k}^{\frac12} (2a)^k n^{-\frac{m}{10}} n^{\frac{m}{1000}}. \] The number of
choices for $\overline g$ and hence $g$ is at most
\[
{n \choose k}^{\frac12} (2a)^k n^{-\frac{m}{10}} n^{\frac{m}{1000}} \cdot k!
\leq {n\choose k}^{\frac12} n^{-\frac{m}{10}} n^{\frac{m}{1000}}(k!)^2
\]
if $m$ is large enough. Hence in this case the number of choices for $g$ is at
most
\[
{n \choose k}^{\frac12} n^{-\frac{m}{10}} n^{\frac{m}{1000}} n^{\frac{m}{50}}
\leq {n \choose k}^{\frac12} \cdot n^{-\frac{m}{20}}.
\]
This completes the proof of the proposition.
\end{proof}

We need the following auxiliary result.
\begin{lem} \label{lem:19} Let $H$ be a permutation group of degree $n$ such
that each element of order $3$ moves at least $100\,000$ points. Assume that $k
\leq n^{\frac23}$ and $k$ is sufficiently large. Then \[ |H_k| \leq {n\choose
k}^{\frac12} (k!)^2. \]
\end{lem}

\begin{proof} Let $g \in H$ be a permutation with $|\text{\rm supp}(g)| = k$
and $|\text{\rm supp}(g_A)| = y$. Using Corollary~\ref{cor:5} we see that the
number of choices for $g_{S\cup L}$ is at most\break ${n \choose k}^{\frac12}
\left[\frac{y}{2}\right]! 2^{(a + 4)k} n^{-\frac{y}{2}} k!$ which is less than
$\frac1k {n \choose k}^{\frac12} (k!)^2 n^{-\frac{y}{2}}$ if $k$ is large
enough. By the remark after Proposition~\ref{prop:8} the number of
possibilities for $g$ is at most $\frac1{k} {n \choose k}^{\frac12} (k!)^2
n^{-\frac{y}{2}} \cdot n^{\frac{y}{5000}} \leq \frac1{k} {n\choose k}^{\frac12}
(k!)^2$. Summing over the $k$ ways to chose $y$, our statement follows.
\end{proof}

\begin{proposition} \label{prop:20}
Assume that $n^{\frac23} \geq k \geq 2^{10\,000}$ and $m$ is sufficiently
large. Then the number of permutations $g$ with $v(g) = v \leq \frac{m}{10}$ is
at most ${n \choose k}^{\frac12} (k!)^{\frac14} n^{-\frac{m}{800\,000\,000}}$.
\end{proposition}

\begin{proof} Just like in the proof of Proposition~\ref{prop:18} we might
assume that $x + y + z \leq \frac{m}{1000}$. Note that in the proof of
Proposition~\ref{prop:18} we do not use the condition on $v$ in the case $k!
\geq n^{\frac{m}{100}}$, so our statement follows in this case. Now assume that
 $k! \leq n^{\frac{m}{100}}$. The number of choices for the $x + y + z + 2v$ points of
$\text{\rm supp}(g)$ which are not contained in the two-element blocks moved by
$g$ is at most $n^{x + y + z + 2v} \leq n^{\frac{m}{5} + \frac{m}{1000}}$. Let
us fix such a set $R$ of $x + y + z + 2v$ points and count the permutations $g$
which correspond to~$R$. Denote by $\mathcal P$ the set of two-element blocks
disjoint from~$R$. Each of the permutations $g$ considered induces a
permutation $\widehat g$ of $\mathcal P$ of support $\frac12 (k - |R|)$. It is
clear that $\text{\rm supp}(\widehat g)$ and $R$ determine $\text{\rm
supp}(g)$. Assume first that $k \geq |\mathcal P|^{\frac23}$. In this case the
number of choices for the two-element blocks moved by  $\widehat g$ is at most
$|\mathcal P|^{\frac{k}{2}} \leq k^{\frac34 k} \leq k!$. Hence the number of
choices for $\text{\rm supp}(g)$ is at most $n^{\frac{m}{4}} \cdot k! \leq
n^{\frac{m}{4} + \frac{m}{100}}$. Applying Theorem \ref{thm:A}, the number of
choices for $g$ itself is bounded by $n^{\frac{m}{4} + \frac{m}{100}}
k^{\frac{k}{10\,000}} \leq \binom{n}{k}^{\frac{1}{2}}$. In this case our
statement follows. Assume now that $k \leq |\mathcal P|^{\frac23}$. Consider
the permutation group $\widehat G$ generated by all the permutations~$\widehat
g$. We claim that each element of order $3$ in $\widehat G$ moves at least
$\frac{m}{4}$ points (of $\mathcal P$). For otherwise let $\widehat h$ be an
element of order $3$ in $\widehat G$ with $|\text{\rm supp}(\widehat h)| \leq
\frac{m}{4}$. Now $\widehat h$ can be written as a product $\widehat h =
\widehat g_1 \dots \widehat g_t$ in $\widehat G$ (where the $\widehat g_i$ are
from the above generating set of $\widehat G$, i.e. each $\widehat g_i$ comes
from one of the $g$). Consider $h = g_1 \dots g_t \in G$. It has order
divisible by $3$ and hence $h^2$ is non-trivial. But $h^2$ moves only points in
$R$ and the points corresponding to the two-element blocks in $\text{\rm
supp}(\widehat h)$. Hence we have $|\text{\rm supp}(h^2)| \leq \frac{m}{2} +
|R| < m$, a contradiction. Applying Lemma~\ref{lem:19}, we see that the number
of possibilities for $\text{\rm supp}(\widehat g)$ is at most $\left({|\mathcal
P| \atop \left[ \frac{k}{2} \right]} \right)^{\frac12} (k!)^2 \leq \left({n
\atop \left[\frac{k}2\right]} \right)^{\frac12} n^{\frac{m}{50}}$ if $m$ is
large enough. Hence the number of choices for $\text{\rm supp}(g)$ is at most
\[ \left({n\atop \left[\frac{k}{2}\right]} \right)^{\frac12} n^{\frac{m}{5} +
\frac{m}{50} + \frac{m}{100}} \leq \left({n \atop \left[\frac{k}{2}\right]}
\right)^{\frac12} n^{\frac{m}{4}}. \] The number of choices for $g$ is at most
$\left({n \atop \left[\frac{k}{2}\right]}\right)^{\frac12} n^{\frac{m}{4}} k!
\leq \left({n \atop \left[\frac{k}{2}\right]} \right)^{\frac12} n^{\frac{m}{4}
+ \frac{m}{100}}$ which implies our statement.
\end{proof}

Putting together Lemma~\ref{lem:2}, Proposition~\ref{prop:18} and
Proposition~\ref{prop:20} we obtain Theorem~\ref{th:size}.

\newpage

\section*{Acknowledgment} JK acknowledges support by the European Commission
under the Integrated Projects RESQ, IST-2001-37559 and Qubit Applications
(QAP) funded by the IST directorate as Contract Number 015848, and by ACI
S\'ecurit\'e Informatique SI/03 511 and ANR AlgoQP grants of the French
Research Ministry. AS wishes acknowledge an Israeli Science Foundation grant
and a Bi-National Science Foundation United States - Israel grant.

\bigskip
\bigskip


\newcommand{\etalchar}[1]{$^{#1}$}

\end{document}